\documentclass[3p,times,procedia]{elsarticle}
\usepackage{nupha_ecrc}

\volume{00}

\firstpage{1}

\journalname{Nuclear Physics A}

\runauth{J.W. Qiu et al.}

\jid{nupha}

\jnltitlelogo{Nuclear Physics A}

\usepackage{amssymb}

\usepackage[figuresright]{rotating}

\usepackage{xcolor}
\usepackage[normalem]{ulem}

\begin{document}

\begin{frontmatter}



\dochead{XXVIIIth International Conference on Ultrarelativistic Nucleus-Nucleus Collisions\\ (Quark Matter 2019)}

\title{QCD factorization and universality of jet cross sections in heavy-ion collisions}

\author[label1]{Jian-Wei Qiu}
\author[label2,label3]{Felix Ringer}
\author[label1]{Nobuo Sato}
\author[label4]{Pia Zurita}

\address[label1]{Theory Center, Jefferson Laboratory, Newport News, Virginia 23606, USA}
\address[label2]{Department of Physics, University of California, Berkeley, CA 94720, USA}
\address[label3]{Nuclear Science Division, Lawrence Berkeley National Laboratory, Berkeley, California 94720, USA}
\address[label4]{Institut f\"ur Theoretische Physik, Universit\"at Regensburg, 93040 Regensburg, Germany}

\begin{abstract}
We review a recently proposed phenomenological framework to establish
the notions of QCD factorization and universality of jet cross
sections in the heavy-ion environment. First results of a global analysis of the nuclear modification factor of inclusive jets are presented where we extract medium modified jet functions using a Monte Carlo sampling approach. We observe that gluon jets are significantly more suppressed than quark jets. In addition, we study the jet radius dependence of the inclusive jet cross section in heavy-ion collisions and comment on a recent measurement from CMS. By considering for example jet substructure observables it will be possible to test the universality of the extracted medium jet functions. We thus expect that the presented results will eventually allow for extractions of medium properties with a reduced model bias.
\end{abstract}


\end{frontmatter}

\section{Introduction}
\label{sec:Introduction}

The quark-gluon plasma (QGP) produced in  heavy-ion collisions at the LHC and RHIC has been conjectured to have filled our universe shortly after the big
bang. 
Highly energetic particles and jets that are also produced in these
collisions serve as important probes of the hot and dense medium. 
These so-called hard probes traverse the QGP and carry information
about the medium properties. 
One of the most striking signatures of the QGP is jet quenching. 
It is often quantified in terms of the nuclear-modification
factor $R_{\rm AA}^{\rm jet}$ which is given by the ratio of the jet
yield in heavy-ion and a rescaled proton-proton baseline. 
One of the main advantages of hard probes is that
the relevant cross sections can be calculated perturbatively in
proton-proton collisions by making use of factorization
theorems~\cite{Collins:1989gx,Nayak:2005rt} which allow for a
consistent separation of perturbative and nonperturbative but
universal ingredients such as parton distribution and fragmentation
functions. 
By analyzing heavy-ion cross sections in terms of leading power
proton-proton factorization theorems, we eventually aim to answer the question of
how much information factorization theorems established in the vacuum tell us
about the corresponding heavy-ion cross sections. 
While it is generally accepted that the main reasons for jet quenching
are parton energy loss and multiple scatterings with spectator partons
in the medium, the exact mechanism remains unknown which can introduce
a significant model bias for extractions of medium properties from
data.

Since there are no first principles proofs of QCD factorization and
universality in heavy-ion collisions, we recently proposed in
Ref.~\cite{Qiu:2019sfj} to establish these concepts
phenomenologically. 
Instead of designing a general purpose parton shower, 
we employ factorization theorems of jet cross sections
which involve jet functions that have been developed for proton-proton
collisions over the last decade. 
If jets are sufficiently collimated, the dynamics of
the formation and evolution of jets can be expressed in terms jet
functions up to corrections which are power suppressed by 
${\cal O}(R^2)$. 
These jet functions are perturbatively calculable in the
vacuum and we perform a first extraction of the analogue
nonperturbative functions in heavy-ion collisions. 
We employ a Monte Carlo sampling technique to reliably extract these
jet functions from the available data. 
Eventually, the universality of the obtained jet functions will have
to be tested by considering other cross sections as well. 
An important example are jet substructure observables where
the extracted medium jet functions are needed to calculate the
relevant quark/gluon fractions~\cite{Ringer:2019rfk}. 
Therefore, the results presented here constitute only a first step in
this direction.

\section{Theoretical framework}

At leading power, the factorized cross sections for inclusive jets
differential in the jet transverse momentum $p_T$ and rapidity $\eta$
in proton proton collisions can be written as
as~\cite{Kaufmann:2015hma,Kang:2016mcy,Dai:2016hzf}
\begin{equation}\label{eq:factorization}
\frac{{\rm d} \sigma^{p p \to {\rm jet}+X}}{{\rm d} p_{T}\, {\rm d} \eta}
  =  \sum_{a b c} f_{a/p} \otimes f_{b/p} \otimes H_{a b}^c \otimes J_c \,,
\end{equation}
where $f_{a,b/p}$ denote the PDFs, $H_{ab}^c$ are hard-scattering
functions of partons $ab\to c$ and the functions $J_c$ take into account the
formation and evolution of the inclusive jet sample originating from
parton $c$. 
The symbols $\otimes$ denote appropriate integrals over
the longitudinal momentum fractions of the involved partons. 
The jet functions $J_c(z,p_T R,\mu)$ depend on the momentum fraction
$z$ of the identified jet relative to the initial parton $c$, the jet
transverse momentum $p_T$, the jet radius $R$ and the renormalization
scale $\mu$. 
The jet functions satisfy the usual timelike DGLAP
evolution equations similar to fragmentation functions
\begin{equation}
\mu\frac{{\rm d}}{{\rm d}\mu}J_c=\sum_d P_{dc} \otimes J_d\,.
\label{eq:dglap}
\end{equation}
For the kinematics considered here, we expect that the only relevant
modification in heavy-ion collisions is the final state jet function.
This assumption is based, for example, on the fact that the photon
yield in heavy-ion collisions is consistent with no modification.
\begin{figure}[!t]
\includegraphics[width=\textwidth]{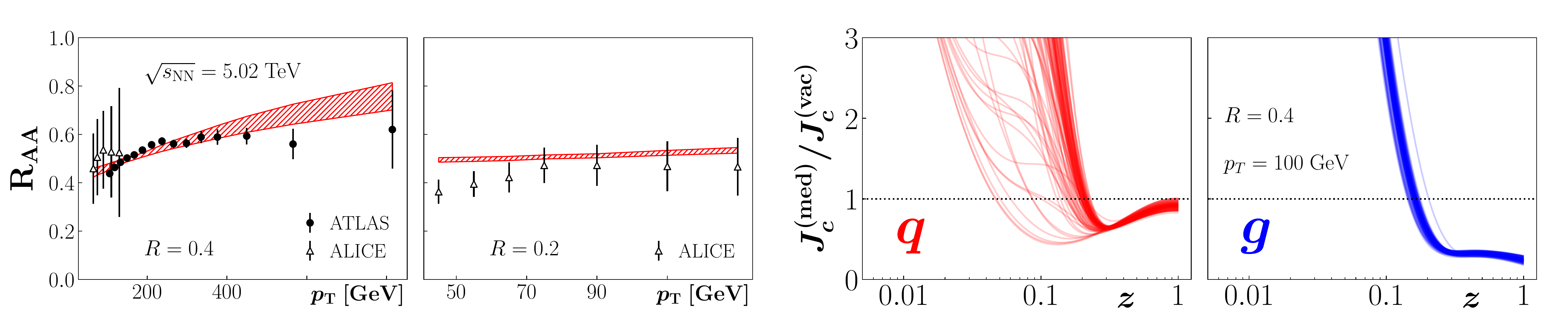}
\caption{Left two panels: The nuclear modification factor $R_{\rm AA}^{\rm jet}$ for inclusive jet production. We show a comparison of our theoretical results using the extracted medium jet functions and LHC data from ATLAS~\cite{Aaboud:2018twu} and
  ALICE~\cite{Acharya:2019jyg}. Right two panels: Ratio of the extracted medium and vacuum jet functions for quarks (red) and gluons (blue).}
\label{fig:1}
\end{figure}
Therefore, we make the following ansatz for the corresponding
heavy-ion cross section where we replace the vacuum jet function in
Eq.~(\ref{eq:factorization}) as
\begin{equation}
    J_c(z,p_T R,\mu)\to J_c^{\rm med}(z,p_T R,\mu) = W_c(z)\otimes J_c(z,p_T R,\mu) \,.
\end{equation}
Here we write the medium jet functions $J_c^{\rm med}$ in terms of the
vacuum ones convolved with weight functions $W_c$ which we fit to
the available data. 
See also~\cite{Kang:2017frl,He:2018gks}. 
We choose a suitable functional form for $W_c$, see~\cite{Qiu:2019sfj}
for more details. 
A related analysis of fragmentation functions in cold nuclear matter
was carried out in~\cite{Sassot:2009sh}. 
We note that in heavy-ion collisions both the initial condition of the
evolution at scale $\sim p_T R$ can be modified as well as the
evolution of the jet functions to the hard scale $\sim p_T$. 
Here we start with the minimal assumption that the jet function
gets affected by energy scales around $\sim p_T R$ which sets the
initial scale for the evolution and constitutes the lowest scale relevant for this process.
We discuss the implications of this assumption in more detail below.

\section{Phenomenological results}

\begin{figure}[!t]
\includegraphics[width=\textwidth]{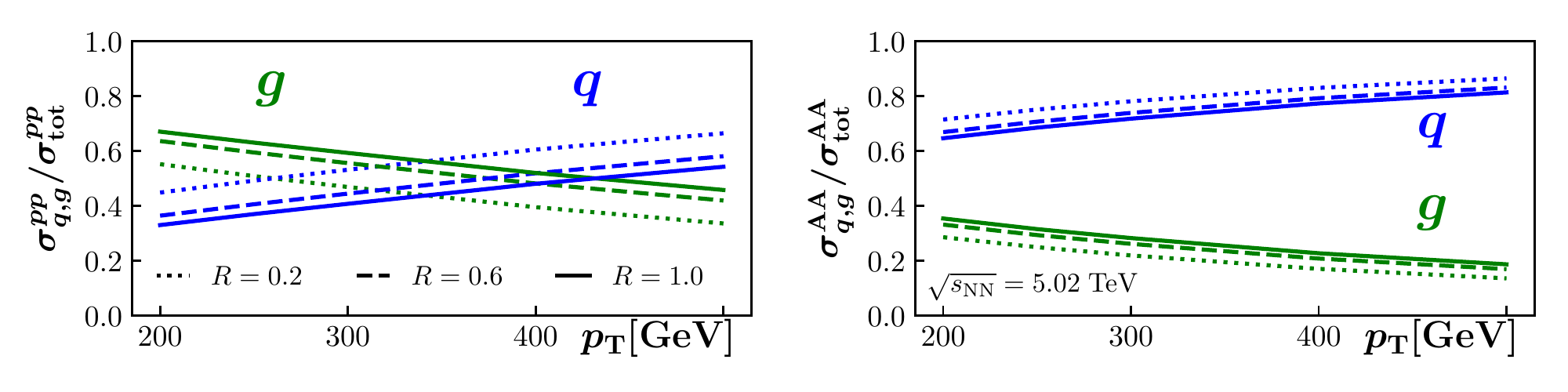}
\caption{The fractions of quark (blue) and gluon (green) jets in proton-proton (left) and heavy-ion collisions (right). We show the results for three values of the jet radius $R$ (different dashing).}
\label{fig:2}
\end{figure}

We consider the available data from 
  ATLAS~\cite{Aaboud:2018twu} and
  ALICE~\cite{Acharya:2019jyg} 
at $\sqrt{s_{\rm NN}}=5.02$~TeV for
central collisions (0-10\%) and $R=0.2,\, 0.4$. 
A similar analysis of the data sets at $\sqrt{s_{\rm NN}}=2.76$~TeV
can be found in~\cite{Qiu:2019sfj}. 
We employ the data resampling technique used in PDF fits such
as~\cite{Ball:2017nwa,Accardi:2016qay}.
In Fig.~\ref{fig:1} (left panel), we present the experimental data and
our theoretical results using the extracted medium jet functions
$J_c^{\rm med}$. 
Overall we observe good agreement and we find $\chi^2/d.o.f=1.7$. 
In the right panel of Fig.~\ref{fig:1}, we show the ratio of the
medium and the vacuum jet functions for $p_T=100$~GeV and $R=0.4$. 
We observe a suppression at large values of $z$ and an
enhancement at small-$z$, which effectively requires the colliding partons to have a larger average momentum fractions $x$ 
in heavy-ion collisions than that in vacuum in order to produce the same $p_T$ jet, 
and leads to 
an overall suppression
of the jet cross section in heavy-ion collisions.
We note that the uncertainty of the extracted jet functions is larger
at small-$z$ which is due to the convolution structure of the cross
section in Eq.~(\ref{eq:factorization}). 
In addition, we observe a more significant modification of the gluon
jet function (right) compared to the quark case (left). 
We explore the strong gluon suppression at the
cross section level by studying the quark/gluon fractions in the
vacuum and medium in more detail~\cite{Banfi:2006hf}.
Fig.~\ref{fig:2} (left) shows the vacuum quark/gluon fractions as a
function of the jet transverse momentum $p_T$ for different values of
the jet radius. 
For smaller jet radii, more evolution makes the jet functions larger at small-$z$, which requires a larger average $x$ and increases the fraction of quark jets.
Similarly, the right panel shows the medium case where we use the extracted medium jet functions for the same center of mass energy.
We observe a significant shift toward quark jets in heavy-ion collisions relative to the vacuum
due to the enhancement of jet functions at small-$z$.
We note that the $R$ dependence of the quark/gluon fractions is reduced in the medium but the ordering compared to the vacuum is preserved.

Recently, CMS published preliminary inclusive jet data for jet radii
in the range of $R=0.2$-1.0~\cite{CMS:2019btm}. 
The experimental data for the ratios of cross sections
with different $R$ is very precise and can provide important
constraints on the extracted medium jet functions discussed here. 
While a more in-depth analysis of the new data set is necessary, we
note that preliminary studies indicate 
that a modification of the DGLAP evolution equations in heavy-ion
collisions could alter the $R$ dependence of the nuclear modification 
factor $R_{\rm AA}^{\rm jet}$.
So far we have assumed that the evolution equations are the same as in the vacuum, see
Eq.~(\ref{eq:dglap}). 
The new data set for different jet radii will thus allow for precision determinations of the medium jet functions and more definitive conclusions about the structure of jet cross sections in heavy-ion collisions can be obtained.

\section{Conclusions and outlook}

In heavy-ion collisions the yield of high transverse momentum jets is
suppressed which is known as jet quenching. 
We employed a factorization ansatz for the jet cross section in
heavy-ion collisions in terms of hard-scattering and medium modified jet functions which
can be determined from data. 
We have presented results of a first global analysis of the
available data from the LHC. 
We observed a significant difference between the suppression of quark
and gluon jets. 
Ultimately, the goal of our analysis is to test or establish the
notions of QCD factorization and universality in heavy-ion collisions. 
We expect that the analysis performed here can eventually lead to
definitive and model independent conclusions about the modification of
the final state parton cascade in heavy-ion collisions and, thus,
provide constraints on models of the QGP and its interaction with hard
probes. 
This can be achieved by extending the presented framework to other
processes such as inclusive hadrons, jet substructure observables and
photon-jet correlations.

\section*{Acknowledgement} 

This work was supported in part by the U.S. Department of Energy under Contract Nos. DE-AC05-06OR23177, DE-AC0205CH11231, the National Science Foundation under Grant No. ACI-1550228 within the JETSCAPE Collaboration and the LDRD program at LBNL. P.Z. was partially supported by the Deutsche Forschungsgemeinschaft (DFG, German Research Foundation) - Research Unit FOR 2926, grant number 409651613.


\bibliographystyle{elsarticle-num}
\bibliography{bibliography}

\end{document}